\documentclass[twocolumn,preprintnumbers,showpacs,showkeys,amsmath,amssymb]{revtex4}
\usepackage{graphicx}
\usepackage{dcolumn}
\usepackage{bm}
\begin{document}
\title{Connection between matrix-product states and superposition of Bernoulli shock measures}
\author{Farhad H. Jafarpour}
\email{farhad@ipm.ir}
\author{Ali Aghamohammadi}
\affiliation{$^1$ Physics Department, Bu-Ali Sina University, Hamadan, Iran}
\date{\today}
\begin{abstract}
We consider a generalized coagulation-decoagulation system on a one-dimensional discrete lattice with reflecting boundaries. It is known that a Bernoulli shock measure with two shock fronts might have a simple random-walk dynamics, provided that some constraints on the microscopic reaction rates of this system are fulfilled. Under these constraints the steady-state of the system can be written as a linear superposition of such shock measures. We show that the coefficients of this expansion can be calculated using the finite-dimensional representation of the quadratic algebra of the system obtained from a matrix-product approach.
\end{abstract}
\maketitle
\section{Introduction}
It is known that the time evolution of a product shock measure in some of the one-dimensional driven-diffusive systems is similar to that of a random-walker, provided that the microscopic reaction rates of the system lie on a certain manifold \cite{KJS}-\cite{AA1}. These systems can be defined both on infinite lattices or finite lattices with boundaries. For each of these systems one can define a microscopic shock position and calculate the exact hopping rates of the traveling wave in terms of the transition rates of the microscopic model. The existence of such processes implies a rather remarkable property. Shocks behave like collective single-particle excitations on the lattice scale which results in reduction of the exponentially large number of microscopic internal degrees of freedom to an only polynomially large number of macroscopically relevant degrees of freedom. Since the shock position moves like a biased single-particle random-walk, the shock measure evolves in time into a linear combination of shock measures; therefore, a linear combination of shock measures may be a stationary measure.

The stationary measure of some of the one-dimensional driven-diffusive systems can also be obtained using a matrix-product approach (for a review see \cite{BE}). According to this approach the stationary probability of a given configuration is written in terms of the expectation value of a product of non-commuting operators associated with different states of each lattice site. In the case of nearest-neighbor interactions these operators should satisfy a quadratic algebra. The matrix representations of some of these quadratic algebras have already been studied in different of cases (see \cite{BE} and references therein).

Since the steady-states of these systems are unique, one can ask about the relation between these two approaches. It has been shown that if the steady-state of a system with nearest-neighbor interactions can be written in terms of superposition of product shock measures with a single shock front, then a two-dimensional representation of the quadratic algebra would be enough to write the steady-state of the system as a matrix-product state \cite{JM1}. In this case the conditions, under which a product shock measure with a single shock front has a random-walk dynamics, are exactly those for the existence of a two-dimensional matrix representation of the quadratic algebra. A couple of examples have already been studied in \cite{JM1}.

In a recent paper \cite{JA} the authors have studied a generalized non-conservative reaction-diffusion system defined on a finite lattice with reflecting boundaries with the following non-zero reaction rates:
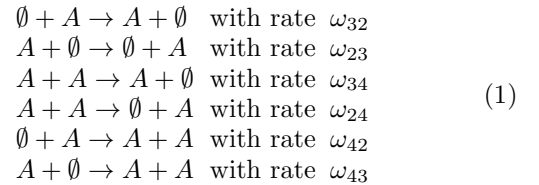
\begin{equation}
\label{Rules}
\begin{array}{ll}
\emptyset+A \rightarrow A+\emptyset & \mbox{with rate} \; \; \omega_{32} \\
A+\emptyset \rightarrow \emptyset+A & \mbox{with rate} \; \; \omega_{23} \\
A+A \rightarrow A+\emptyset & \mbox{with rate} \; \; \omega_{34} \\
A+A \rightarrow \emptyset+A & \mbox{with rate} \; \; \omega_{24} \\
\emptyset+A \rightarrow A+A & \mbox{with rate} \; \; \omega_{42} \\
A+\emptyset \rightarrow A+A & \mbox{with rate} \; \; \omega_{43} \\
\end{array}
\end{equation}
in which $A$ and $\emptyset$ stand for the presence of a particle and a hole in each lattice site respectively. Note that there is no injection or extraction from the boundaries. It has been shown that the steady-state of this system can be written as a matrix-product state under some constraints \cite{JA}. The quadratic algebra of the system has a four-dimensional representation in this case. The time evolution of a product shock measure with two shock fronts has also been studied for this model. It has been shown that the shock positions in this measure have simple random-walk dynamics provided that the \emph{same} conditions for the existence of the four-dimensional matrix representation for the quadratic algebra of the system are fulfilled by the microscopic reaction rates \cite{JA}. In other words these shock distributions form an invariant sector under the time evolution of the system i.e. a shock measure evolves into a linear combination of shock measures with different shock positions. In this case one can write the steady-state of the system as a linear superposition of shocks; however, this has not been shown yet. The complete phase diagram of this system has also been studied. It has been found that it has two different phases depending on the hopping rates of the shock positions.

In this paper we aim to study the steady-state of the exactly solvable system defined in (\ref{Rules}) using a different approach than the matrix-product approach. We will show that the steady-state of the system can be expand in terms of product shock measures with two shock fronts. We will specifically show that the coefficients of the expansion can be uniquely calculated using the four-dimensional matrix representation of the quadratic algebra introduced in \cite{JA}. In the following we will start with the mathematical preliminaries and bring the time evolution equations for a product shock measure with two shock fronts generated by the Hamiltonian of the system. By considering a linear superposition of such shock measures we construct the steady-state of the system and then calculate the coefficients of this expansion. We will provide the summery and discussion at the end of the paper.

\section{Temporal evolution of shocks}

The microscopic state of a one-dimensional reaction-diffusion system may be described by a set of occupation numbers $\{n_1,\cdots,n_L\}$ where $n_k = 0,1$ is the number of particles on site k on a lattice of L sites. The time evolution of the distribution $P(\{n_1,\cdots,n_L\}; t )$ in this system is defined by a continuous-time master equation which can be written in terms of the quantum Hamiltonian formalism. The stochastic Hamiltonian $H$, whose matrix elements are the transition rates between configurations, generates the time evolution. The Markovian time evolution can be written in the form of an imaginary time Schr\"odinger equation
\begin{equation}
\frac{d}{dt} \vert P(t )\rangle= H \vert P(t )\rangle.
\end{equation}
For a single-species system with nearest-neighbors interactions defined on a one-dimensional lattice of length $L$ with reflecting boundaries the Hamiltonian $H$ is of the following form:
\begin{equation}
\label{H1}
H=\sum_{k=1}^{L-1}h_{k,k+1}
\end{equation}
in which:
$$
h_{k,k+1}={\mathcal I}^{\otimes (k-1)}\otimes h \otimes {\mathcal I}^{\otimes (L-k-1)}
$$
where ${\mathcal I}$ is a $2 \times 2$ identity matrix and $h$ is a $4
\times 4$ matrix representing the bulk interactions. In the basis $(00,01,10,11)$ with the following basis vectors:
\begin{equation}
\label{Basis}
 \vert n_k=1 \rangle =
  \left(\begin{array}{c}
    0 \\ 1
  \end{array}\right), \;
  \vert n_k=0\rangle =
  \left(\begin{array}{c}
    1 \\ 0
  \end{array}\right)\; \mbox{for $k=1,\cdots,L$}
\end{equation}
the Hamiltonian $h$ for the system defined by (\ref{Rules}) can be written as:
\begin{equation}
\begin{array}{c}
\label{H2}
h = \left( \begin{array}{cccc}
0 & 0                        &             0 & 0 \\
0 & -\omega_{32}-\omega_{42} & \omega_{23} & \omega_{24} \\
0 & \omega_{32}              & -\omega_{23}-\omega_{43} & \omega_{34} \\
0 & \omega_{42}              & \omega_{43} & -\omega_{34}-\omega_{24} \end{array} \right).
\end{array}
\end{equation}
In \cite{JA} the authors have shown that a product shock measure with two shock fronts defined as:
\begin{equation}
\label{Shock measure}
  \vert P_{m,n} \rangle =
  \left(\begin{array}{c}
    1 \\ 0
  \end{array}\right)^{\otimes m} \otimes
  \left(\begin{array}{c}
    1-\rho \\ \rho
  \end{array}\right)^{\otimes n-m-1}
\otimes
  \left(\begin{array}{c}
    1 \\ 0
\end{array}\right)^{\otimes L-n+1}
\end{equation}
for $0 \leq m \leq n-1$ and $1 \leq n \leq L+1$ might evolve according to 2-particle random-walk dynamics provided that:
\begin{equation}
\label{Constraints}
\frac{1-\rho}{\rho}:=\frac{\omega_{24}+\omega_{34}}{\omega_{42}+\omega_{43}}=
\frac{\omega_{23}}{\omega_{43}}=\frac{\omega_{32}}{\omega_{42}}
\end{equation}
Note that in (\ref{Shock measure}) two auxiliary sites $0$ and $L+1$ are defined for convenience. The time evolution of the product shock measure (\ref{Shock measure}) is obtained to be:
\begin{widetext}
\begin{equation}
\label{Time Evoulution}
\begin{array}{l}
H \vert P_{m,n} \rangle = \delta_{1r} \vert P_{m+1,n} \rangle + \delta_{1l} \vert P_{m-1,n} \rangle
+\delta_{2r}\vert P_{m,n+1} \rangle+\delta_{2l} \vert P_{m,n-1}\rangle -(\delta_{1r}+\delta_{1l}+\delta_{2r}+\delta_{2l})\vert P_{m,n} \rangle \\
\mbox{for} \quad m=1,\cdots,L-2 \quad \mbox{and}\quad n=m+2,\cdots,L \\ \\
H \vert P_{0,n} \rangle=-\bar\delta \vert P_{1,n} \rangle+\delta_{2r}
\vert P_{0,n+1} \rangle+\delta_{2l} \vert P_{0,n-1} \rangle
-(-\bar\delta+\delta_{2r}+\delta_{2l})\vert P_{0,n} \rangle \quad \mbox{for} \quad n=2,\cdots,L\\ \\
H \vert P_{m,L+1} \rangle=\delta_{1r} \vert P_{m+1,L+1} \rangle+\delta_{1l} \vert
P_{m-1,L+1} \rangle+\bar\delta\vert P_{m,L} \rangle-
(\delta_{1r}+\delta_{1l} +\bar\delta)\vert P_{m,L+1} \rangle \quad \mbox{for} \quad m=1,\cdots,L-1\\ \\
H\vert P_{0,L+1} \rangle=-\bar\delta\vert P_{1,L+1} \rangle+\bar\delta\vert P_{0,L} \rangle\\ \\
H\vert P_{m,m+1} \rangle=0 \quad \mbox{for} \quad m=0,\cdots,L
\end{array}
\end{equation}
\end{widetext}
in which we have defined:
\begin{equation}
\label{Ref}
\begin{array}{l}
\delta_{1l}:=\frac{\omega_{42}}{\rho}, \\
\delta_{1r}:=\omega_{43}\frac{(1-\rho)^2}{\rho}+\omega_{24}\rho, \\
\delta_{2l}:=\omega_{42}\frac{(1-\rho)^2}{\rho}+\omega_{34}\rho, \\
\delta_{2r}:=\frac{\omega_{43}}{\rho}\\
\bar\delta:=\delta_{2l}-\frac{\delta_{1r}+\delta_{2l}}{\delta_{1l}+\delta_{2r}}\delta_{2r}.
\end{array}
\end{equation}
As can be seen in the bulk of the lattice the shock positions move like two biased lattice random-walks; therefore, provided that we are far from the boundaries, the velocity, and also the diffusion coefficient of each shock front, can be easily calculated from (\ref{Time Evoulution}). The hopping rates of the left (right) shock front are $\delta_{1r},\delta_{1l}$ ($\delta_{2r},\delta_{2l}$). The velocity of an individual shock front is then $v_i=\delta_{ir}-\delta_{il}$ for $i=1,2$ and also the diffusion coefficient is $D_i=(\delta_{ir}+\delta_{il})/2$
for $i=1,2$.

\section{Steady-State of the system}

The last equation in (\ref{Time Evoulution}) implies that an empty lattice is a trivial steady-state for the system; however, one can construct a non-trivial steady-state by considering a linear superposition of $\vert P_{m,n}\rangle$'s as follows:
\begin{equation}
\label{steady-state}
\vert P^{\ast} \rangle =\frac{1}{Z_{L}} \sum_{m=0}^{L}\sum_{n=m+1}^{L+1} \psi_{m,n}\vert P_{m,n} \rangle
\end{equation}
and find the coefficients $\psi_{m,n}$'s by requiring that:
\begin{equation}
\label{steady-state Condition}
H \vert P^{\ast} \rangle=0.
\end{equation}
The probability of finding the system in the configuration $\{n_1,\cdots,n_L\}$ is now given by:
\begin{equation}
\label{Probaility}
P^{\ast}(\{n_1,\cdots,n_L\})=(\langle n_1 \vert \otimes\cdots \otimes \langle n_L\vert) \vert P^{\ast} \rangle
\end{equation}
in which $\vert n_k \rangle$ for $k=1,\cdots,L$ is defined in (\ref{Basis}). Note that there are $L+1$ states $\vert P_{k,k+1}\rangle$'s for $k=0,\cdots,L$ which all point to an empty lattice and should be considered as a single state. The coefficient of this state in (\ref{steady-state}) will be called $\psi'$.

It is a lengthy exercise, but nevertheless straight-forward to find the equations governing $\psi_{m,n}$'s using
(\ref{Time Evoulution}), (\ref{steady-state}) and (\ref{steady-state Condition}). It turns out that these coefficients should satisfy a set of fifteen equations which are given in Appendix. In order to find the non-trivial steady-state of the system one should solve these equations and calculate the coefficients $\psi_{m,n}$'s. These equations have already been solved in \cite{JM2} for the following special tuning of the parameters:
\begin{equation}
\label{Conditions}
\begin{array}{c}
\omega_{24}=\omega_{23}=q^{-1}\; , \; \omega_{34}=\omega_{32}=q\; , \\ \\
\omega_{42}=\Delta q \;,\;\omega_{43}=\Delta q^{-1}.
\end{array}
\end{equation}
In this paper we are not going to solve these equations directly, but instead, we will find their solutions using the results obtained from the matrix-product approach.

The non-trivial steady-state of the system has already been obtained using the matrix-product approach \cite{JA}. Let us assign the two operators $D$ and $E$ to the existence of a particle and a vacancy at each lattice site respectively. According to this approach the normalized matrix-product steady-state is given by:
\begin{equation}
\label{matrix-product Weight}
\vert P^{\ast} \rangle=\frac{1}{Z_{L}} \langle\langle W \vert
\left(
  \begin{array}{c}
    E \\
    D \\
  \end{array}
\right)^{\otimes L}\vert V \rangle\rangle
\end{equation}
in which $Z_L=\langle\langle W \vert(D+E)^L\vert V \rangle\rangle$. In contrast to \cite{JA}, here we use a new notation for the two vectors $\vert V \rangle\rangle$ and $\langle \langle W \vert$ to emphasis that the vectors in the configuration space denoted by $\vert \cdots \rangle$ are different from these two vectors which exist in an auxiliary space. Using (\ref{Probaility}) the probability of finding the system in the configuration $\{n_1,\cdots,n_L\}$ is now given by:
$$
P^{\ast}(\{n_1,\cdots,n_L\})=\frac{1}{Z_{L}}\langle\langle W \vert \prod_{k=1}^{L}(n_k D+(1-n_k)E) \vert V \rangle\rangle.
$$
In \cite{JA} the authors have found that the two operators $D$ and $E$ besides the two vectors $\vert V \rangle\rangle$ and $\langle\langle W \vert$ have a four-dimensional matrix representation provided that the constraints (\ref{Constraints}) are satisfied. Because of uniqueness of the steady-state of the system, the expression (\ref{steady-state}) should be equal to the expression (\ref{matrix-product Weight}). This will help us calculate the coefficients $\psi_{m,n}$'s. It can easily be checked that the results obtained in this way satisfy the equations governing $\psi_{m,n}$'s.

\section{Equivalence of two different approaches}

Let us define a new measure:
\begin{equation}
\label{New measure}
  \vert \tilde P_{m,n} \rangle =
  \left(\begin{array}{c}
    1 \\ \frac{1-\rho}{-\rho}
  \end{array}\right)^{\otimes m} \otimes
  \left(\begin{array}{c}
    0 \\ \frac{1}{\rho}
  \end{array}\right)^{\otimes n-m-1}
\otimes
  \left(\begin{array}{c}
    1 \\ \frac{1-\rho}{-\rho}
\end{array}\right)^{\otimes L-n+1}
\end{equation}
for $0 \leq m \leq n-1$ and $1 \leq n \leq L+1$ which is orthogonal to the product shock measure (\ref{Shock measure}) according to the following rule:
\begin{equation}
\langle \tilde P_{m',n'}\vert P_{m,n} \rangle =\delta_{m,m'}\delta_{n,n'}.
\end{equation}
Using this new measure and (\ref{steady-state}) one finds:
\begin{equation}
\label{Psi}
\psi_{m,n}=Z_{L}\langle \tilde P_{m,n}\vert P^{\ast} \rangle
\end{equation}
for $0 \leq m \leq n-1$ and $1 \leq n \leq L+1$ which uniquely determines the coefficients $\psi_{m,n}$'s. Since the two steady-states obtained from the superposition of the product measures and the one obtained from the matrix-product approach should be equal, one finds from (\ref{matrix-product Weight}), (\ref{New measure}) and (\ref{Psi}):
\begin{widetext}
\begin{equation}
\label{Psi MPF}
\begin{array}{l}
\psi_{m,n}=\langle\langle W \vert (E-\frac{1-\rho}{\rho}D)^{m}(\frac{1}{\rho}D)^{n-m-1}(E-\frac{1-\rho}{\rho}D)^{L-n+1}\vert V \rangle\rangle \; \mbox{for} \; 0 \leq m \leq L-1 \; \mbox{and} \;  m+2 \leq n \leq L+1, \\ \\
\psi'=\langle\langle W \vert (E-\frac{1-\rho}{\rho}D)^{L}\vert V \rangle\rangle.
\end{array}
\end{equation}
\end{widetext}
If there exists a matrix representation for the quadratic algebra of the system, one can calculate the coefficients $\psi_{m,n}$'s using (\ref{Psi MPF}), and it means that the matrix-product steady-state of the system
can actually be expand in terms of a linear superposition of product shock measures with two shock fronts.

Using the diagonal four-dimensional matrix representation of the quadratic algebra of the system, first introduced in \cite{JA}, we have calculated the coefficients $\psi_{m,n}$'s. It turns out that these unnormalized coefficients are given by the following expression:
\begin{widetext}
\begin{equation}
\label{Psi sol}
\begin{array}{l}
\psi_{m,n}=\frac{(\frac{\delta_{1r}(\delta_{1l}+\delta_{2r})}
{\delta_{1r}\delta_{2r}-\delta_{1l}\delta_{2l}})^{\delta_{m,0}}}
{(\frac{\delta_{1l}\delta_{2l}-\delta_{1r}\delta_{2r}}{\delta_{2l}(\delta_{1l}+\delta_{2r})})^{\delta_{n,L+1}}}
(\frac{\delta_{1r}}{\delta_{1l}})^m(\frac{\delta_{2r}}{\delta_{2l}})^n
((\frac{\delta_{1r}+\delta_{2l}}{\delta_{1l}+\delta_{2r}})^{n-m}-
(\frac{\delta_{1r}+\delta_{2l}}{\delta_{1l}+\delta_{2r}}))  \;
\mbox{for} \; 0 \leq m \leq L-1 \; \mbox{and} \; m+2 \leq n \leq L+1,\\ \\
\psi'= \frac{\rho\delta_{1l} \delta_{1r}}{(\delta_{1l}+\delta_{2r})}(\frac{\delta_{2r}(
\delta_{2l}+\delta_{1r})}{\delta_{1l} \delta_{2l}-\delta_{1r} \delta_{2r}})^2(\frac{(
\delta_{2l}+\delta_{1r})\rho}{(\delta_{1r}-\delta_{1l}(1-\rho)^2)(\delta_{2l}-
\delta_{2r}(1-\rho)^2)}(\frac{\delta_{2r}}{\delta_{2l}}(1-\rho)^2)^L -\frac{1}
{(\delta_{1r}-\delta_{1l}(1-\rho )^2)}(\frac{\delta_{1r} \delta_{2r}}{\delta_{1l} \delta_{2l}})^L-\frac{1}{(\delta_{2l}-\delta_{2r}(1-\rho )^2)})
\end{array}
\end{equation}
\end{widetext}
in which $\rho=1-(\delta_{1r}+\delta_{2l})/(\delta_{1l}+\delta_{2r})$. These coefficients are functions of the four parameters $\delta_{1r},\;\delta_{1l},\;\delta_{2r}$ and $\delta_{2l}$ as one should expect. As we mentioned it can be easily verified that $\psi_{m,n}$'s besides $\psi'$ in (\ref{Psi sol}) satisfy the equations (\ref{Coef Equ}). Note that in order to calculate the coefficient $\psi'$ in (\ref{Psi sol}) one should also include the fact that an empty lattice (the trivial steady-state) should be excluded by applying the constraint $P^{\ast}(\{0,\cdots,0\}) \propto \langle\langle  W \vert E^L \vert V \rangle\rangle=0$. One can also check that for the special tuning of the parameters given in (\ref{Conditions}), the unnormalized coefficients (\ref{Psi sol}) reduce to the ones calculated in \cite{JM2} (up to a multiplicative constant).

As we mentioned the system has two different phases which are separated by a coexistence line. On this coexistence line, where one of the shock fronts performs an unbiased random-walk, the matrix representation of the quadratic algebra cannot be diagonalized and one should use a non-diagonal matrix representation. It can be easily shown that even in this case one can calculated $\psi_{m,n}$'s and $\psi'$ using (\ref{Psi MPF}) and the non-diagonal matrix representation of the quadratic algebra introduced in \cite{JA}.

\section{Conclusion}

In this paper we have studied the steady-state of a generalized coagulation-decoagulation system defined on a finite lattice with reflecting boundaries. The steady-state of this system has already been found using a matrix-product approach; however, in this paper we have shown that the steady-state of the system can equivalently be constructed by considering a linear superposition of product shock measures with two shock fronts. The key point is that the conditions under which the shock fronts have simple random-walk dynamics, are exactly those necessary for the existence of a four-dimensional representation for the quadratic algebra of the system. This is quite non-trivial since it has not been generally shown that the existence of a finite-dimensional matrix representation for the quadratic algebra of a given system is an indication that the steady-state of the system can be written as a superposition of Bernoulli shock measures. The results of the study of this exactly solvable system besides those obtained in \cite{JM1} might bring us to this conclusion: that the existence of a finite-dimensional matrix representation for the algebra of a one-dimensional reaction-diffusion system is a signal that the steady-state of the system can be written as a superposition of product shock measures; nevertheless, the proof still remains as an open problem.

\appendix
\section{Equations governing $\psi_{m,n}$'s}
As we mentioned, the non-trivial steady-state of the system can be written as a linear superposition of Bernoulli shock measures with two shock fronts of type (\ref{Shock measure}). By requiring that (\ref{steady-state}) is the
steady-state of the system and using (\ref{Time Evoulution}), (\ref{steady-state}) and (\ref{steady-state Condition}) one finds that the coefficients of this expansion should satisfy the following difference equations:
\begin{widetext}
\begin{equation}
\label{Coef Equ}
\begin{array}{l}
(\bar\delta-\delta_{2l})\psi_{0,2}-(\delta_{1r}+\delta_{2l})\sum_{m=1}^{L-2}\psi_{m,m+2}-(\bar\delta+\delta_{1r})
\psi_{L-1,L+1}=0 \\
(\bar\delta-(\delta_{2r}+\delta_{2l}))\psi_{0,2}+\delta_{2l}\psi_{0,3}=0 \\
\delta_{2r}\psi_{0,n-1}+(\bar\delta-(\delta_{2r}+\delta_{2l})\psi_{0,n}+\delta_{2l}\psi_{0,n+1}+
\delta_{1l}\psi_{1,n}=0 \quad\mbox{for $n \neq 2,L$} \\
\bar\delta\psi_{0,L+1}+\delta_{2r}\psi_{0,L-1}+(\bar\delta-(\delta_{2r}+\delta_{2l}))\psi_{0,L}+
\delta_{1l}\psi_{1,L}=0 \\
\delta_{2r}\psi_{0,L}+\delta_{1l}\psi_{1,L+1}=0 \\
\bar\delta\psi_{0,3}+(\delta_{1r}+\delta_{1l}+\delta_{2r}+\delta_{2l})\psi_{1,3}-\delta_{2l}\psi_{1,4}=0 \\
\bar\delta\psi_{0,n}-\delta_{2r}\psi_{1,n-1}+(\delta_{1r}+\delta_{1l}+\delta_{2r}+\delta_{2l})\psi_{1,n}-\delta_{2l}
\psi_{1,n+1}-\delta_{1l}\psi_{2,n}=0 \quad \mbox{for $n \neq 3,L$} \\
\bar\delta(\psi_{0,L}-\psi_{1,L+1})-\delta_{2r}\psi_{1,L-1}+(\delta_{1r}+\delta_{1l}+\delta_{2r}+\delta_{2l})\psi_{1,L}
-\delta_{1l}\psi_{2,L}=0 \\
\bar\delta\psi_{0,L+1}+(\bar\delta+\delta_{1r}+\delta_{1l})\psi_{1,L+1}-\delta_{1l}\psi_{2,L+1}-\delta_{2r}\psi_{1,L}=0\\
\delta_{1r}\psi_{m-1,L+1}-(\bar\delta+\delta_{1r}+\delta_{1l})\psi_{m,L+1}+\delta_{1l}\psi_{m+1,L+1}+
\delta_{2r}\psi_{m,L}=0 \quad \mbox{for $m \neq 1,L-1$} \\
\delta_{1r}\psi_{L-2,L+1}-(\bar\delta+\delta_{1r}+\delta_{1l})\psi_{L-1,L+1}=0 \\
\bar\delta\psi_{m,L+1}+\delta_{1r}\psi_{m-1,L}+\delta_{2r}\psi_{m,L-1}-(\delta_{1r}+\delta_{1l}+\delta_{2r}+\delta_{2l})
\psi_{m,L}+\delta_{1l}\psi_{m+1,L}=0 \quad \mbox{for $m \neq 0,1,L-1,L-2$} \\
\bar\delta\psi_{L-2,L+1}+\delta_{1r}\psi_{L-3,L}-(\delta_{1r}+\delta_{1l}+\delta_{2r}+\delta_{2l})\psi_{L-2,L}=0 \\
\delta_{1r}\psi_{m-1,n}-(\delta_{1r}+\delta_{1l}+\delta_{2r}+\delta_{2l})\psi_{m,n}+\delta_{2l}\psi_{m,n+1}=0 \quad
\mbox{for $m \neq 0,1,L-2,L-1$ and $n=m+2$}.\\
\delta_{1r}\psi_{m-1,n}+\delta_{2r}\psi_{m,n-1}-(\delta_{1r}+\delta_{1l}+\delta_{2r}+\delta_{2l})\psi_{m,n}+
\delta_{2l}\psi_{m,n+1}+\delta_{1l}\psi_{m+1,n}=0 \\
\mbox{for $m \neq 0,1,L-1$ and $n \neq m+2,L,L+1$}
\end{array}
\end{equation}
\end{widetext}
Note that $\bar{\delta}$ is related to the shock hopping rates through the last relation in (\ref{Ref}).

\end{document}